# How to identify earth pressures on in-service tunnel linings: Insights from Bayesian inversion to address non-uniqueness

Zhiyao Tian[1,2,3], Shunhua Zhou[1,2*], Anthony Lee[3], Yao Shan[1,2], Bettina Detmann[4]

[1] *Key Laboratory of Road and Traffic Engineering of the Ministry of Education, Tongji University, Shanghai, China*

[2] *Shanghai Key Laboratory of Rail Infrastructure Durability and System Safety, Tongji University, Shanghai, China*

[3] *School of Mathematics, University of Bristol, Bristol, United Kingdom*

[4] *Department of Civil Engineering, Faculty of Engineering, University of Duisburg-Essen, Essen, Germany*

*\*Corresponding author, E-mail address: zhoushh@tongji.edu.cn. (Shunhua Zhou)*

**Abstract:** Identifying earth pressures on in-service transportation tunnel linings is essential for their health monitoring and performance prediction, particularly in structures that exhibit poor performance. Due to the high costs associated with pressure gauges, pressure inversion based on easily observed structural responses, such as deformations, is preferred. A significant challenge lies in the non-uniqueness of inversion results, where various pressures can yield similar structural responses. Existing approaches often overlook detailed discussions on this critical issue. In addressing this gap, this study introduces a Bayesian approach. The proposed statistical framework effectively quantifies the uncertainty induced by non-uniqueness. Further analysis identifies the uniform component in distributed pressures as the primary source of non-uniqueness. Insights into mitigation strategies are provided, including increasing the quantity of deformation data or incorporating an observation of internal normal force within the tunnel lining — the latter proving to be notably more effective. A practical application in a numerical case study demonstrates the effectiveness of this approach. In addition, our investigation recommends maintaining deformation measurement accuracy within the range of [−1, 1] mm to ensure satisfactory outcomes. Finally, deficiencies and potential future extensions of this approach are discussed.

**Keywords:** Earth pressures; Tunnel linings; Load inversion; Non-uniqueness; Bayesian approach

# 1 Introduction

Global transportation demands have significantly increased the use of tunnel linings, especially in subway systems [1]. Nevertheless, the maintenance and health monitoring of a high volume of in-service structures present considerable challenges for engineers. Empirical evidence has demonstrated a concerning trend: numerous newly-built tunnels have exhibited substantial deterioration in their structural performance [2–4].

A primary cause for this deterioration could be the notable changes in external earth pressures on the linings. Specifically, disturbances from the operational environment can induce pressures far exceeding those anticipated during the design phase. For example, a section of Nanjing Metro Line 3 encountered disruptions due to nearby excavation activities, resulting in severe distortions and notable defects in the linings, thereby threatening the metro's operational safety [5]. Similar instances have been globally reported, particularly in cities with soft soil conditions [6–9].

For these poorly performing (resulting from environmental disturbance) in-service tunnel linings, identifying the current earth pressures—the primary trigger—is critical for effective health monitoring and performance prediction. However, directly measuring the external loads via pressure gauges often proves challenging [10–12]. Specifically, since the in-service linings have already been cast and constructed, embedding sensors within the concrete is impractical. Moreover, even if it were feasible, the extensive deployment of sensors in the field is economically unsustainable. Alternatively, inversion of the earth pressures based on easily observed structural responses, say deformation [13–15], is more desirable.

As presented in Eq. (1), a forward model calculates the structural response under a given load condition:

$$\mathbf{d} = g(\mathbf{q}), \qquad (1)$$

where, $\mathbf{d}$ is the vector of structural responses, such as deformation. For illustration purposes, we define the arbitrary distributed pressure acting on the linings as $\mathcal{F}$ and denote **q** as the vector of parameters representing $\mathcal{F}$; The function $g(.)$ serves as the forward mapping. Conversely, inverse problems estimate the pressure parameters given a set of observed structural responses **d**. A direct approach would entail the inversion of Eq. (1), as shown in Eq. (2):

$$\mathbf{q} = \mathbf{g}^{-1}(\mathbf{d}). \tag{2}$$

However, this inverse mapping is typically intractable due to the non-explicit nature of $g(.)$ [16]. To address this, an optimization framework is often employed, as outlined in Eq. (3):

$$\mathbf{q}_o = \arg\min_{\mathbf{q}}\{\|\mathbf{d}-\mathbf{g}(\mathbf{q})\|_2^2\}, \tag{3}$$

where $\mathbf{q}_o$ denotes the optimized load parameters that yield structural responses bested fitted with the observed data **d**; $\|.\|_2$ represents the second-order norm. Simplifying the inversion process typically involves parameterizing the unknown $\mathcal{F}$ based on a predetermined design mode [17–18], as shown in Fig. 1(a), thereby restricting the unknown parameters to a finite set, such as $\mathbf{q}=(q_1, q_2, q_3, q_4)^\mathrm{T}$.

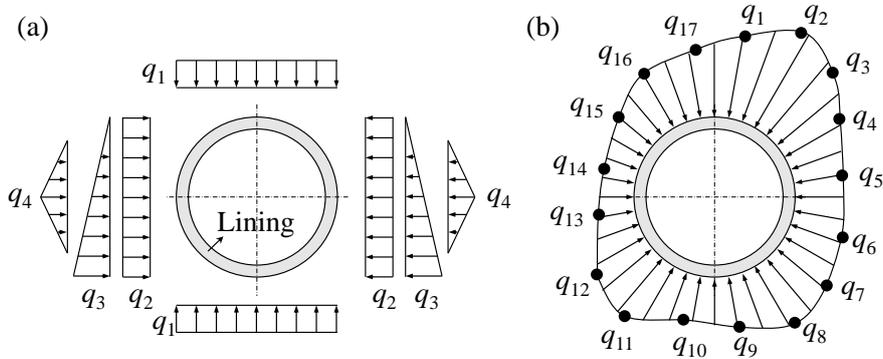

**Fig. 1.** Parameterization strategies for the unknown earth pressures acting on the tunnel linings: (a) parameterization based on a predefined design mode; (b) parameterization utilizing an interpolation technique.

It is important to note, however, that the actual earth pressures exerted on the poorly performing linings may deviate from conventional design modes, often exhibiting uneven distributions, particularly in soft soil conditions [19–21]. For *a priori* unknown pressure distributions, Gioda and Jurina [21] introduced an interpolation technique for parameterization. Specifically, as exemplified in Fig. 1(b), a set of unknow knots was placed within the structural domain, for example $\mathbf{q}=(q_1,\ldots,q_{17})^{\mathrm{T}}$. Interpolation through these points approximates the actual distributed pressures, transforming the inversion of $\mathcal{F}$ into inversion of the unknown interpolation points $\mathbf{q}$.

Gioda and Jurina's method [21] abandons the design mode assumption, which relaxes the inversion parameter space. Without stringent restrictions, two significant issues typically arise: non-uniqueness and ill-conditioning [16,22]. That is, vastly different solutions can yield structural responses fitting equally well with the observed data, and slight errors in the observations can lead to an unstable and fluctuating solution. Similarly, Lin et al. [23] introduced an inversion approach based on Betti's theorem, but seldom considered measurement errors in their validation cases. Liu et al. [24] and Liu et al. [25] have noted the ill-conditioning issue arising from measurement errors. They employed regularization techniques to penalize undesired components in the parameter space, leading to satisfactory and smooth solutions in their cases. However, it is worth noting that all the above-mentioned references conducted inversion within a deterministic framework, and the non-uniqueness of solutions has not been extensively investigated or discussed.

The authors [26] have previously introduced a Bayesian approach for the inversion of earth pressures on in-service underground structures, which is capable of quantifying the non-uniqueness of solutions. While this methodology has been applied to structures such as diaphragm walls and anti-slide piles, it has not yet been extended to tunnel linings. This paper aims to fill this gap by adapting this method to the distinct characteristics of tunnel linings. Most importantly, by quantifying non-uniqueness via this method, this study will reveal the inherent complexities—specifically, the sources of non-uniqueness in inversion results—that are unique to tunnel linings and not encountered with other structures. Furthermore, this paper outlines strategies for mitigating non-uniqueness and achieving improved inversion results. This presented approach and corresponding findings may provide valuable insights for the digital maintenance of tunnel linings.

## 2 Methods

This approach comprises three major steps: i) a parameterization process to represent the *a priori* unknown distributed pressure; ii) a Bayesian framework to mathematically describe the posterior distribution of the unknown parameters; and iii) a sampling process to estimate the posterior distribution of the unknown pressures.

### 2.1 *Parameterization of the distributed pressures*

Two assumptions are made before parameterization: i) the load inversion in this study focuses on 2-D problems (i.e., adopting a plane strain assumption), which draws insight from the long and linear characteristics of tunnels [27]; ii) only the normal pressures are considered, i.e., the tangential pressures on the lining are assumed to be negligible due to their minimum impact on deformation [25].

We adopted the interpolation technique [21] for the parameterization. As shown in Fig. 2, this method facilitates the approximation of the *a priori* unknown pressure $\mathcal{F}$ through a distribution function $q(\theta)$, where $\theta$ represents the polar coordinate encircling the tunnel lining over a full 360° span. The function $q(\theta)$ is interpolated by a series of evenly distributed knots **q** as outlined in Fig. 2 and Eq. (4):

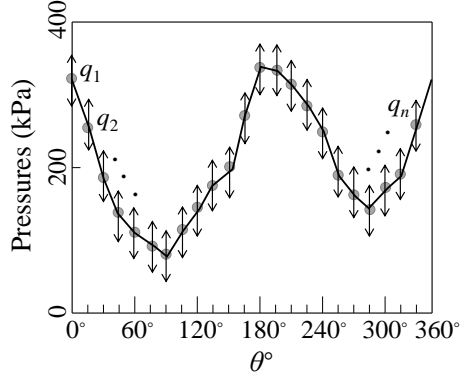

**Fig. 2.** The evenly distributed knots and the interpolated function for parameterization. (Note: this figure presents a planar view similar to Fig. 1(b), unfolded from 0° to 360°)

$$\mathcal{F} \approx q(\theta) = \mathbf{I}(\theta)\mathbf{q}, \qquad (4)$$

where, $\mathbf{I}(\theta)$ is a linear interpolation operator matrix, and its detailed expression can be found in Press et al. [28]. **q** is a vector comprising *n* unknowns, $\mathbf{q}=(q_1,\ldots,q_n)^\mathrm{T}$. As indicated in Fig. 2, the shape of q($\theta$) is controlled by **q**. In this way, the inversion of $\mathcal{F}$ has been transformed into the inversion of the representative parameters **q**.

It is worth noting that the parameterization capability of **q** is determined by its length $n$. That is, the more parameters, the greater the capacity of the interpolation function to fit any shape of distributed pressures. In engineering practice, $n$ can be determined according to engineering judgement. Alternatively, it can be determined by a trial testing method introduced by Tian et al. [26], which conducts trial tests to monitor the inversion results by gradually increasing $n$. Once the results do not change with an increase in $n$, the parameterization capability can be regarded as sufficient.

## 2.2 The Bayesian framework

The inversion of **q** in a Bayesian framework is performed using conditional probability given observed data based on Bayes' rule:

$$p(\mathbf{q}|\mathbf{d}) = \frac{p(\mathbf{q})p(\mathbf{d}|\mathbf{q})}{p(\mathbf{d})}, \qquad (5)$$

where, $p(\mathbf{q}|\mathbf{d})$ is the posterior probability density of **q** given the filed data **d**; $p(\mathbf{q})$ is the prior probability density of **q**; $p(\mathbf{d}|\mathbf{q})$ is the likelihood function; and $p(\mathbf{d})$ is generally referred to as the "marginal likelihood", a normalizing factor that will get canceled in the numerical estimation process of Eq. (5).

### 2.2.1 The prior distribution

The prior, $p(\mathbf{q})$, reflects one's pre-judgement on the values of the parameters beforehand. To avoid imposing constraints on the tuning of **q**, the parameters are assumed to be independent in the prior, yielding:

$$p(\mathbf{q}) = \prod_{i=1}^{n} p(q_i). \qquad (6)$$

In general, making assumptions about the values of the parameters beforehand can be challenging. Thus, flat priors are typically used for the parameters. Specifically, we adopted a uniform prior to indicate that all values for $q_i$ within a physically plausible bound have equal probabilities before inversion:

$$p(q_i) = \begin{cases} \dfrac{1}{q_{\max} - q_{\min}}, & q_{\min} \leq q_i \leq q_{\max} \\ 0, & \text{else} \end{cases}. \tag{7}$$

The weakly informative priors make it suitable for unevenly distributed and *a priori* unknown earth pressures. The physically plausible bound [$q_{\min}$, $q_{\max}$] can be determined based on engineering experience. For example, $q_{\min}$ can be set as 0 since almost no traction can be exerted by the soil, and $q_{\max}$ can be determined to be a very large value according to the specific engineering application. The following illustrative example will provide a reference for determining the bounds.

*2.2.2 The likelihood function*

The likelihood function, $p(\mathbf{d}|\mathbf{q})$, updates the weakly informative priors by incorporating the observed structural response. This function assesses the probability that the parameters $\mathbf{q}$ results in observation $\mathbf{d}$, as quantified by the error vector $\mathbf{e}$:

$$\mathbf{e} = \mathbf{d} - g(\mathbf{q}), \tag{8}$$

where, $\mathbf{d}$ is the observed structural response, such as deformations; $g(.)$ is the forward model mapping any pressure parameters $\mathbf{q}$ to the predicted structural responses, as detailed in section 2.2.3. The error $\mathbf{e}$ arises from inaccuracies in the forward model (model errors) and measurement errors in the observed data. In line with the Central Limit Theorem, it is generally reasonable to assume that $\mathbf{e}$ follows an independent zero-mean Gaussian distribution, denoted as $\mathbf{e} \sim \mathcal{N}(0, \sigma_e^2 \mathbf{I})$, where $\sigma_e$ is the estimated standard deviation of the errors, and $\mathbf{I}$ is the identity matrix.

It is important to note that the output of $g(\mathbf{q})$ given $\mathbf{q}$, denoted as $g(\mathbf{q})|\mathbf{q}$, is a deterministic event, as $g(.)$ is constructed as a deterministic model. Consequently, the distribution of $\{g(\mathbf{q})+\mathbf{e}|\mathbf{q}\}$, equaling to $\mathbf{d}|\mathbf{q}$, is Gaussian centered at $g(\mathbf{q})$, yielding:

$$\mathbf{d} \mid \mathbf{q} \sim \mathcal{N}(g(\mathbf{q}), \sigma_e^2 \mathbf{I}). \tag{9}$$

Thus, the likelihood function is expressed as:

$$p(\mathbf{d}|\mathbf{q}) = \dfrac{1}{(2\pi\sigma_e^2)^{H/2}} \exp\left(-\dfrac{[\mathbf{d}-g(\mathbf{q})]^T [\mathbf{d}-g(\mathbf{q})]}{2\sigma_e^2}\right) \tag{10}$$

where, $H$ is the quantity of observed data, i.e., the lengths of **e** and **d**; In straightforward scenarios involving a highly accurate forward model, it is generally feasible to consider model errors as negligible compared to measurement errors, leading to $\sigma_e$ being predominantly determined by the precision of measurement instruments. It is noteworthy that these assumptions, although introduced for simplicity and practicality in engineering applications, have been empirically validated as effective in numerous instances [22, 26]. Further discussions will also be provided in section 4.3.3.

*2.2.3 The forward model*

As outlined in section 2.2.2, the forward model plays a crucial role in constructing the likelihood function, which predicts structural responses under specified pressures. This model is commonly referred to as the load-structure model in tunnel engineering. Analytical models, such as the force method, typically necessitate the analytical integration of external pressures to establish equilibrium equations [29–30]. However, this integration becomes intractable when pressures are distributed unevenly and irregularly. Consequently, the finite element method (FEM) is advocated in this context, as it facilitates the discretization of pressures, thereby enabling a numerical solution:

$$g(\mathbf{q}) = \mathbf{K}^{-1}\mathbf{f}(\mathbf{I}(\theta)\mathbf{q}), \tag{11}$$

where, the structure is discretized into a series of elements, with **K** representing the global stiffness matrix; **f** being a vector-valued function where $\mathbf{f}(\mathbf{I}(\theta)\mathbf{q})$ is the equivalent nodal forces equivalent to the distributed pressures $\mathbf{I}(\theta)\mathbf{q}$, i.e., $q(\theta)$, following the transformation rules of virtual work. The forward model is determined based on specific case requirements. For the case studies presented in this paper, we adopted the widely used "embedded beam spring model" with its detailed derivation of **K** and **f** provided in Appendix A.

As illustrated in Fig. 3(a), the segments are simulated using Euler beams, incorporating axial stiffness ($EA$) and bending stiffness ($EI$). The beams are connected with a rotation spring characterized by a rotation stiffness ($k_\varphi$). The mechanical behavior of the beams and joints should be determined on a case-by-case basis. It is noteworthy that the generalized framework introduced in this study facilitates adaptation to various scenarios by simply substituting the forward model, thus obviating the need for any modifications to the Bayesian framework itself. For example, compression and shear springs can be incorporated into the forward model by simply replacing Eq. (11), without affecting the inversion process. In the case of linear elasticity, Eq. (11) can be computed directly, whereas an iterative procedure becomes necessary when accounting for nonlinear behavior.

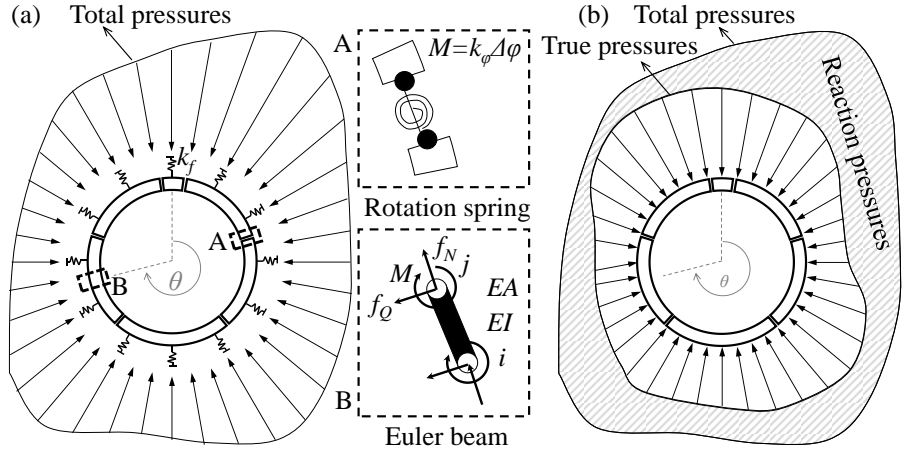

**Fig. 3.** The forward model: (a) the embedded beam spring model; (b) a strategy to address the asymmetric earth pressures.

Different from other underground structures, the boundary conditions for a tunnel lining remain unclear prior to inversion [26, 31]. Specifically, as shown in Fig. 3(a), the earth pressures on the linings can become asymmetric during the inversion process, potentially leading to rigid body displacement and rendering Eq. (11) unsolvable. To address this issue, it is necessary to consider the actual soil-structure interaction and introduce soil springs (characterized by normal stiffness $k_f$) on the linings to counterbalance the asymmetric component of pressures. This involves initially inverting the total earth pressures on the beam-foundation (soil springs) structure, recording the soil reaction pressures (exemplified as shaded areas in Fig. 3b), and then combining the total pressures with the reaction pressures to obtain the true pressures acting directly on the linings. The effectiveness of this strategy will be thoroughly tested and discussed in section 4.3.2.

## 2.3 Estimation of the posterior distribution

Due to the complexity of Eq. (11), finding an analytical solution for Eq. (5) is challenging. Consequently, the Markov Chain Monte Carlo method (MCMC) is employed to estimate the posterior distribution. The fundamental idea behind MCMC involves: i) Constructing a transition kernel, $P(t \rightarrow t+1)$, based on Detailed Balance [32]; ii) Iteratively sampling with $P(t \rightarrow t+1)$ to construct an ergodic Markov Chain. The stationary distribution of this chain converges to the target posterior distribution.

To enhance efficiency, we adopt a variant of MCMC known as the Differential Evolution Markov Chain (DE-MC) [33]. In the transition kernel of DE-MC, $T$ parallel "chains" (components) run simultaneously and "learn" from each other, thereby accelerating the convergence of the chain. The process is detailed as follows:

i) Generate a proposal of the $i$-th chain ($i=1,\ldots,T$) at the iteration step $t$ using the rule of Differential Evolution:

$$\mathbf{q}_p^i = \mathbf{q}_t^i + \lambda(\mathbf{q}_t^a - \mathbf{q}_p^b) + \zeta_n, \tag{12}$$

where $\lambda$ is the jump rate; $\zeta_n \sim \mathcal{N}_n(0,c)$ is drawn from a normal distribution with a small standard deviation to ensure ergodicity; $a$ and $b$ are integer values drawn without replacement from set $\{1,\ldots,i-1,i+1,\ldots,T\}$.

ii) Accept the proposal, $\mathbf{q}_{t+1}^i = \mathbf{q}_p^i$, with probability $p_{acc}(\mathbf{q}_t^i \rightarrow \mathbf{q}_p^i)$; and reject otherwise: $\mathbf{q}_{t+1}^i = \mathbf{q}_t^i$, where

$$p_{acc}(\mathbf{q}_t^i \rightarrow \mathbf{q}_p^i) = \min[1, \frac{p(\mathbf{q}_p^i | \mathbf{d})}{p(\mathbf{q}_t^i | \mathbf{d})}], \tag{13}$$

where the ratio between $p(\mathbf{q}_p^i|\mathbf{d})$ and $p(\mathbf{q}_t^i|\mathbf{d})$ is estimated using Eq. (5). Following these two steps iteratively, the Markov chain is constructed. After the chain convergence, collect the posterior samples $\{\mathbf{q}_s, s=1,\ldots,S\}$ to estimate the posterior distribution:

$$p(\mathbf{q}|\mathbf{d}) \approx \frac{1}{S}\sum_{s=1}^{S}\delta(\mathbf{q}-\mathbf{q}_s), \tag{14}$$

where, $\delta(\cdot)$ is the Dirac delta function. The distribution of $q(\theta)$ can also be estimated by

combining Eq. (4):

$$p(q(\theta)|\mathbf{d}) \approx \frac{1}{S}\sum_{s=1}^{S}\delta(q(\theta)-\mathbf{I}(\theta)\mathbf{q}_s). \tag{15}$$

The posterior means integrates across all individual solutions according to the posterior distribution, serving as a representative solution:

$$E(q(\theta)|\mathbf{d}) = \int \mathbf{I}(\theta)\mathbf{q}p(\mathbf{q}|\mathbf{d})d\mathbf{q}. \tag{16}$$

The posterior means minimize the expected squared loss, flattening the ill-conditioning features of individual solutions [34–36]. Substituting the posterior distribution with Eq. (14), Eq. (16) reduces to:

$$E(q(\theta)|\mathbf{d}) = \frac{1}{S}\sum_{s=1}^{S}\mathbf{I}(\theta)\mathbf{q}_s. \tag{17}$$

The convergence of the Markov components can be monitored by the scale-reduction factor proposed by Gelman and Rubin [37]:

$$\hat{R}_i = \sqrt{\frac{t_i-1}{t_i}+\frac{N+1}{Nt_i}\frac{B}{W}} \tag{18}$$

where, $t_i$ is the number of iterations in the $i$-th chain ($i=1,…,T$); $T$ is the number of chains; $B$ is the variance between chain means; $W$ is the average of within-chain variances; and a recommended convergence criterion for the chain is a value of $\hat{R}_i$ less than 1.2 [33].

## 3 Insights learned from the Bayesian approach

### 3.1 An illustration example

An illustration example is presented to demonstrate the application of the Bayesian approach. Most importantly, this section primarily analyzes what can we learn from this approach by quantifying uncertainty.

### 3.1.1 Preliminaries

As shown in Fig. 4(a), predefined pressures (adapted from Gioda and Jurina [21]) are applied to a tunnel lining. The structural and soil properties are defined as linear-elastic, with parameters detailed in this figure. Using these pressures, we run the forward model (Eq. 11) to calculate the noise-free deformation profile of this structure, as shown in Fig. 4(c). We define 100 evenly distributed baselines, as displayed in Fig. 4(c). Subsequently, we observe the convergence deformations on these baselines, as presented in Fig. 4(d). The objective is then the inversion of the pressures (assumed to be unknown currently) based on these synthetic convergence data. It is worth noting that, in this example, earth pressures are directly applied to the beam-foundation structure, and the inversion target is the total pressure on the beam-foundation structure. Consequently, the process of eliminating the reaction force of soil springs is unnecessary in this particular example.

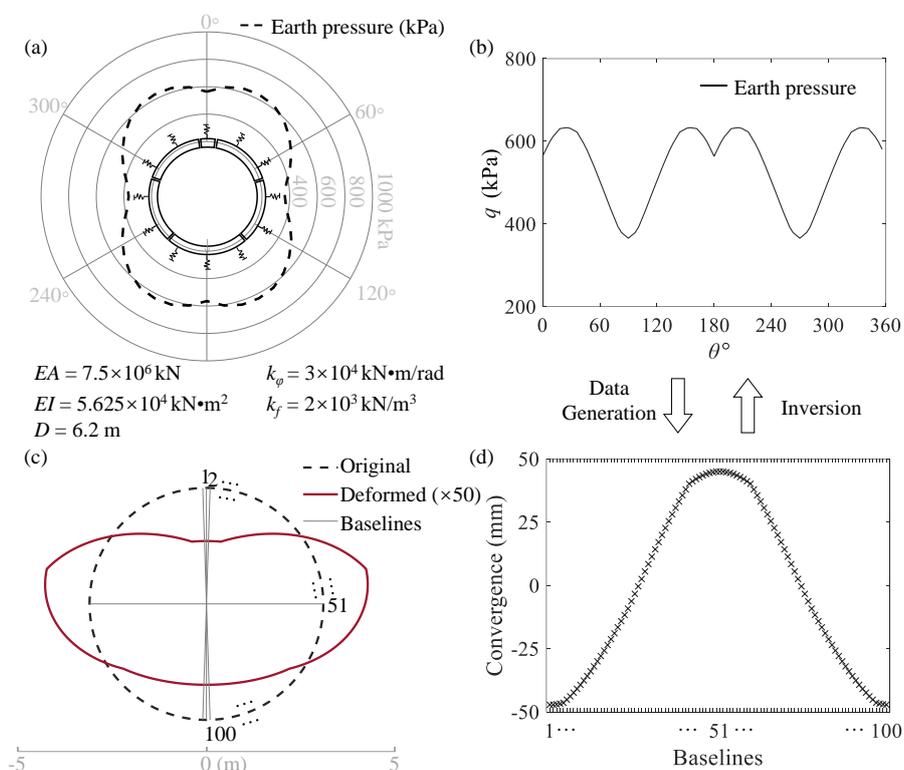

**Fig. 4.** The illustration example: (a) earth pressures adapted from Gioda and Jurina [21]; (b) a planar view of the pressures from 0° to 360°; (c) the calculated deformed profile of the tunnel lining; (d) the noise-free convergence deformation on the defiend baselines. (note: *EA* and *EI* are the axial and bending stiffness of the tunnel segments, respectively; $k_\varphi$ is the rotation stiffness of the joint; $k_f$ is the normal stiffness of the soil springs; and *D* is the diameter of the tunnel)

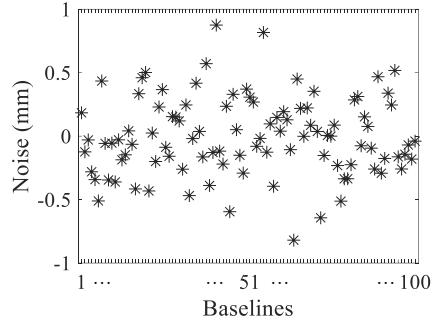

**Fig. 5.** The noise to simulate measurement errors.

To simulate measurement errors commonly encountered in engineering practice, we incorporated Gaussian noise $e_i$, following a distribution $\mathcal{N}(0, 1/3^2)$, into the noise-free convergence data $d_i'$, for $i = 1,\ldots,H$. This distribution aims to constrain the noise within the range of $[-1, 1]$ mm, with a more detailed discussion provided in Section 4.3.1. Fig. 5 illustrates the added noise. Consequently, the actual observations, contaminated by noise and utilized for pressure inversion, are denoted as $\mathbf{d} = (d_1,\ldots,d_H)^{\mathrm{T}}$, where $d_i = d_i' + e_i$, for $i = 1,\ldots,H$.

For the purpose of necessary comparisons, six different cases involving a varying number of convergence data were examined. Specifically, cases A to F comprise 2, 5, 10, 25, 50, and 100, evenly distributed convergence data, respectively. For instance, in the case with only 2 data points, convergence data is available only on baselines 1 and 51 (Fig. 4c) for inversion, while in the case with 100 data points, observations from all 100 baselines are available for inversion.

We employed the index of agreement to evaluate how well the inversion results fit the actual pressures [38]:

$$\mathrm{IA} = 1 - \frac{\sum_{j=1}^{M_p}[q_I(\theta_j) - q_A(\theta_j)]^2}{\sum_{j=1}^{M_p}[|q_I(\theta_j) - \bar{q}_A| + |q_A(\theta_j) - \bar{q}_A|]^2}, \qquad (19)$$

where IA is the index of agreement; $q_I(\theta_j)$ and $q_A(\theta_j)$ denote the inversion and actual pressures at the monitoring point $\theta_j$, respectively; 100 monitoring points are distributed evenly across the structural domain spanning from 0° to 360°, and $M_p$ is the quantity of the monitoring points, i.e., $M_p$ =100; $\bar{q}_A$ is the mean of the actual pressures at all monitoring points. The index of agreement, IA ranges from 0 to 1, and the closer IA is to 1, the closer the inversion result aligns with the actual values. While not strictly rooted in statistical principles, we follow the recommendation from previous studies, and set a threshold of 0.7 [39]. Values below this threshold indicate poor agreement between inversion results and observed data. In addition, the RMSE (Root Mean Square Error) is employed to measure the average magnitude of the errors between inversion and actual pressures:

$$\text{RMSE} = \sqrt{\frac{\sum_{j=1}^{M_p}[q_I(\theta_j)-q_A(\theta_j)]^2}{M_p}}. \tag{20}$$

Besides, a standard deviation factor Std is introduced to assess the uncertainty of the inversion results, which averages the standard deviations across all monitoring points:

$$\text{Std} = \frac{1}{M_p}\sum_{j=1}^{M_p}\sigma_{\theta_j}, \tag{21}$$

where, $\sigma_{\theta j}$ is the standard deviation of the posterior distribution at monitoring point $\theta_j$.

*3.1.2 Inversion process and parameter settings*

Twenty-two unknown parameters are set evenly distributed across the structural domain, drawing insights from previous experience [26]. These parameters, denoted as **q**=$(q_1, \ldots, q_{22})^T$, are set approximately every 16° from 0° to 360°. The determination of the parameter quantity (density) will be discussed following the analysis of the results. The prior distribution of these parameters is set as a uniform distribution. Based on engineering judgement, the lower bound is set as 0 kPa. The upper bound is set as 3000 kPa, a sufficiently loose bound in engineering practice. Consequently, a uniform prior is set as $q_i \sim$ Uniform (0,3000) ($i$=1,…,22). The prior distribution of earth pressures along the tunnel lining from 0° to 360° is presented in Fig. 6. The flat-distributed colormap indicates equal probabilities of any values within (0, 3000) kPa in the structural domain. Additionally, the mean values of the prior distribution equals 1500 kPa, which does not align with the true pressures and warrants an update via information from the likelihood function.

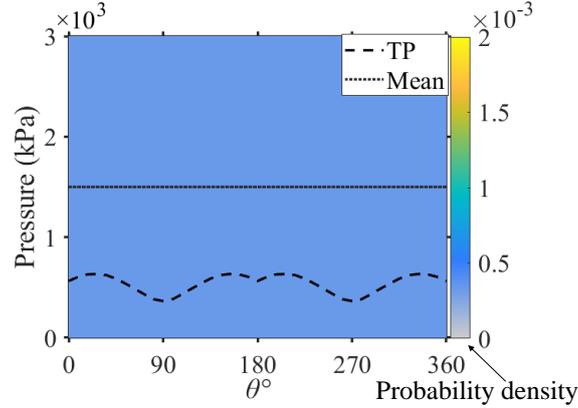

**Fig. 6.** The prior distribution of earth pressures along the tunnel lining from 0° to 360°. (Note: TP= true pressures)

According to the precision of measurement accuracy of convergence deformation in engineering practice, $\sigma_e$ in the likelihood function is determined to be 1 mm. The forward model in the likelihood function is constructed with the process illustrated in section 2.2.3 and the parameters presented in Fig. 4. The sampling steps introduced in Section 2.3 are used to build the Markov Chains. Following Ter Braak et al. [33], the number of components $T$ should be at least $2n$ (where $n$ is number of the parameters). As a result, $T$ is taken to be 44, and the number of iterations is set to be 20000. For comparison, an accompanying deterministic inversion (Eq. 3) is also run for every case, respectively, using the same input data and equivalent bounded constraints within [0, 3000] kPa [26].

*3.1.3 Results*

Take case A as an example to present the sampling process. As seen in Fig. 7, the scaling-reduction factors of the 22 parameters $\hat{R}_i\ (i=1,\ldots,22)$ rapidly converged to values below the defined threshold of 1.2 at approximately 4000 steps, indicating that the components approach their stationary distribution. Consequently, the last 50% of the samples in the chains are used to estimate the posterior distribution.

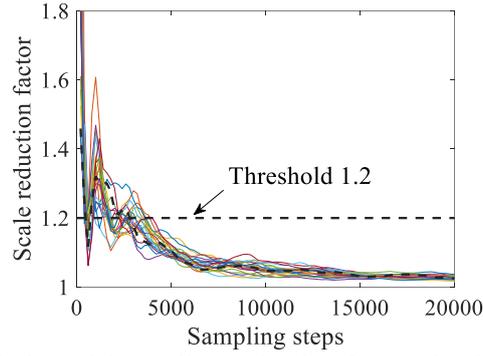

**Fig. 7.** Evolution of the scaling-reduction factors for the parameters.

The inversion results for cases A–F are presented in Figs. 8(a)–(f), respectively. The estimated probability density of pressures at various angles within the structural domain is illustrated by various color levels quantitatively. Lighter colors correspond to higher probabilities, while darker colors denote lower probabilities. Notably, as observation data are incorporated, the posterior distribution begins to take shape, distinguishing itself from the initially flat prior distribution (Fig. 6). As more data are involved, uncertainty in the posterior distribution diminishes, with lighter colors increasingly converging towards the true pressures (TP). Simultaneously, the posterior means (PM) become progressively more aligned with TP.

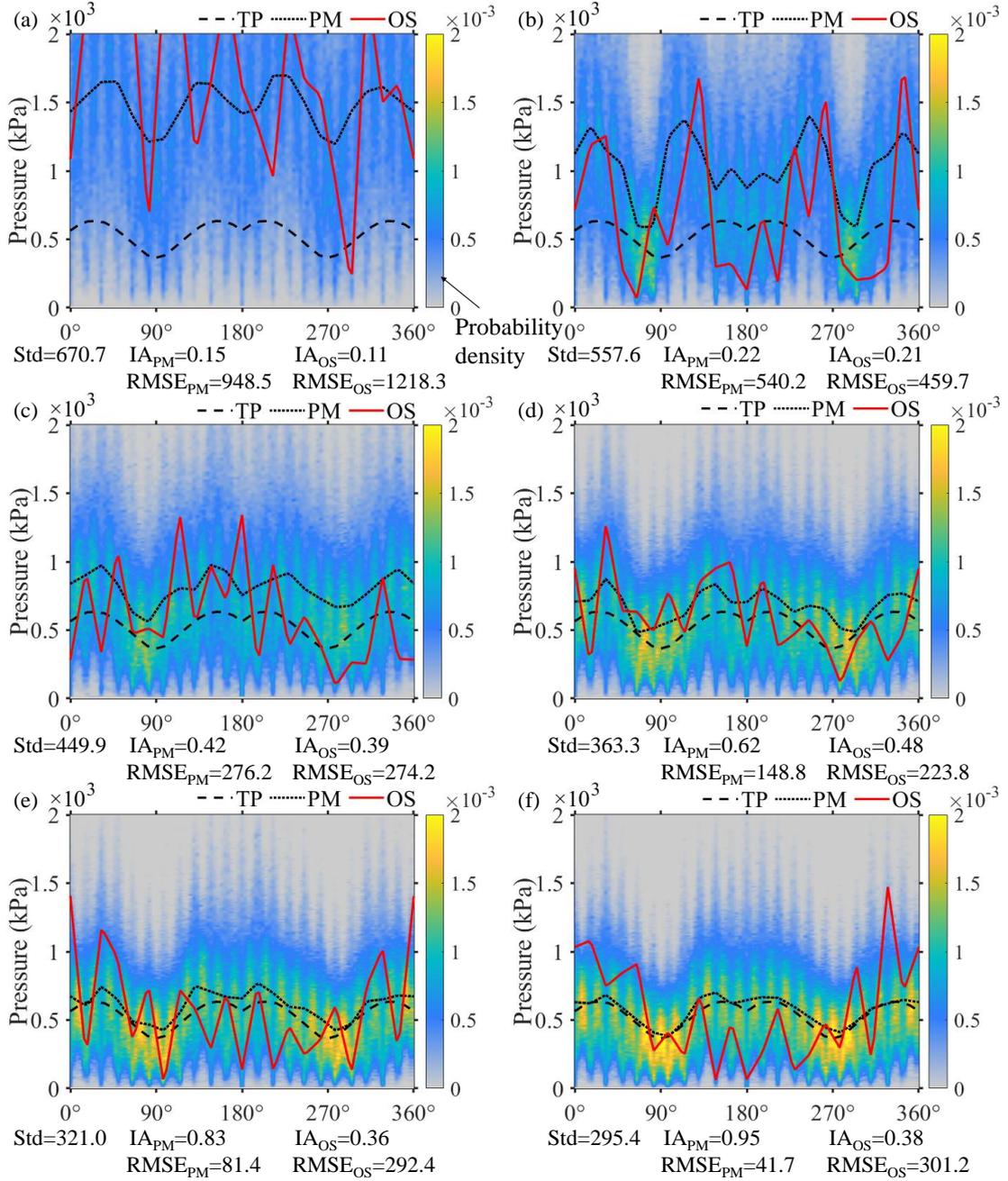

**Fig. 8.** Inversion results for: (a) case A (2 convergence data); (b) case B (5 convergence data); (c) Case C (10 convergence data); (d) case D (25 convergence data); (e) case E (50 convergence data); (f) case F (100 convergence data). (TP= true pressures; PM=posterior mean; OS=optimal solution obtained by deterministic inversion; the y-axis are rescaled from 0 to 2000 kPa to provide a clearer result)

It is worth noting that PM serves as a representative solution of the overall posterior distribution, exhibiting much smoother than the optimal solution (OS) obtained by deterministic inversion (Eq. 3). In contrast, the OS encounters ill-conditioning, which fluctuated dramatically in all the 6 cases, resulting in low values for the index of agreement $IA_{OS}$ and high values for $RMSE_{OS}$. These results highlight the PM's capacity to mitigate ill-conditioning. A comprehensive discussion and theoretical underpinnings of this capacity can be found in the references [26,35].

Of paramount significance, it should be noted that PM effectively fit with the true pressures (TP) only when the input data quantity reaches 50 and 100 (cases E and F) with $IA_{PM}$=0.83 and 0.95, $RMSE_{PM}$=81.4 and 41.7 kPa, respectively. Interestingly, for cases A–D, although $IA_{PM}$ for PM fitting with TP is less than the threshold 0.7, the shapes of the curves closely resemble those of TP. It seems as though PM in cases A–F can be considered as shifted versions of the true values, and the role of adding deformation data is to reduce uncertainty and "push" PM towards the true pressures. This observation may be critical in understanding the origin of uncertainty, and will be discussed in detail in the following sections.

*3.2   Where does the uncertainty arise from?*

Generally, uncertainty in the inversion results arises from two aspect: random measurement errors and the non-uniqueness of the mechanical system itself. While we will discuss the former in Section 4, we focus on the latter in this section.

As shown in Fig. 9(a), two different pressures, namely $q_1$ and $q_2$, are applied to a tunnel lining, while Fig. 9(b) presents an unfolded view from 0° to 360°. Notably, $q_2$ is generated by adding a uniform pressure of 200 kPa to $q_1$, and visually, $q_2$ can be considered a translation of $q_1$. Using $q_1$ and $q_2$, we run the forward model to calculate the deformation of the lining, and observe convergence data on the 100 baselines as shown in Fig. 10. The deformation dataset $d_1'$ is generated by $q_1$, while $d_2'$ is generated by $q_2$. It is evident that $d_1'$ almost coincides with $d_2'$, with only minor differences, as indicated in the zoomed-in box.

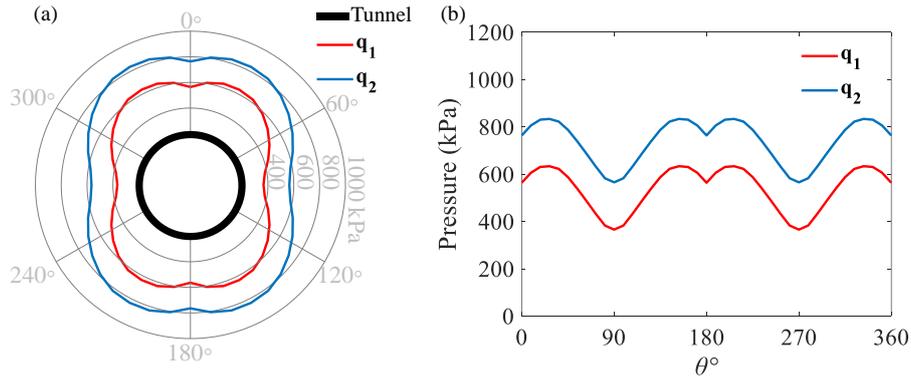

**Fig. 9.** Applying two set of pressures on a tunnel lining: (a) $q_1$ and $q_2$ within the structural domain from 0° to 360°; (b) a planar view.

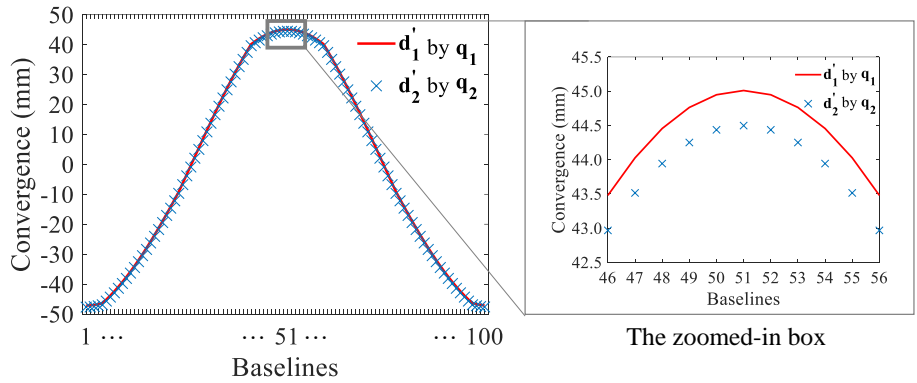

**Fig. 10.** The deformations yielded by $q_1$ and $q_2$

From a different perspective, the difference between $q_1$ and $q_2$ is the uniform pressures. As illustrated in Fig. 11(a), when uniform pressures are applied to a tunnel lining, it results in uniform convergence across the structural domain. This convergence is generally extremely low due to the high value of the axial stiffness *EA*. Fig. 11(b) indicates that the convergence is only 0.5 mm at all baselines when the uniform pressures are set at 200 kPa, which can easily be obscured by measurement noise. This leads to the minor differences observed between $d_1'$ and $d_2'$. These minor differences ultimately introduce non-uniqueness and high uncertainty in the inversion process. That is, vastly different pressures, say $q_1$ and $q_2$, can yield deformation data that fit equally well with the observed data.

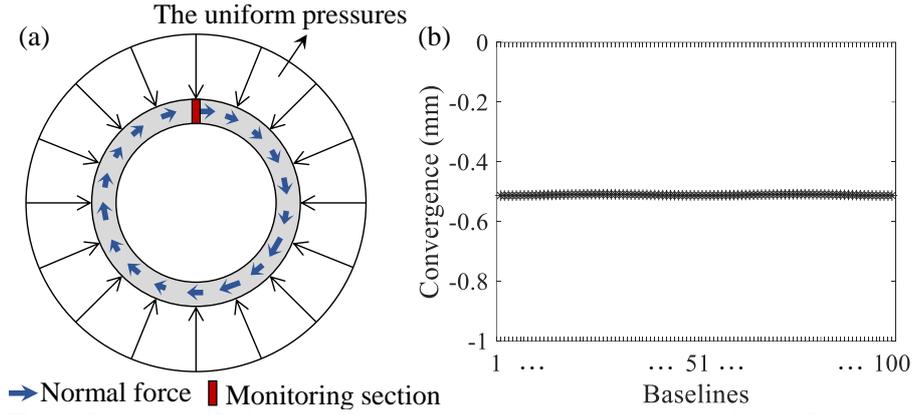

**Fig. 11.** The deformations yielded by uniform pressures on a tunnel: (a) the uniform pressures of 200 kPa; (b) the convergence deformation yielded by the uniform pressures

Mathematically, as demonstrated in Eq. (5), the posterior probability density of $\mathbf{q}_1$ and $\mathbf{q}_2$, given observed deformations, is determined by both the prior distribution and the likelihood function. Since the priors for $\mathbf{q}_1$ and $\mathbf{q}_2$ are identical (Fig. 6), the critical factor distinguishing $\mathbf{q}_1$ from $\mathbf{q}_2$ is the likelihood function. For simplification, the logarithmic likelihood for $\mathbf{q}_1$ and $\mathbf{q}_2$ is given as follows:

$$\log \mathcal{L}_j \propto \sum_{i=1}^{H} (d'_{i,j} - d_i)^2 \quad j=1,2, \tag{22}$$

where, $d'_{i,j}$ is the deformation data at the $i$-th baseline yielded by earth pressures $\mathbf{q}_j$, $j=1$ and 2, respectively; $d_i$ is the actual observed deformation data at the $i$-th baseline; $H$ is the quantity of observed deformation data. By taking the difference between $\log \mathcal{L}_1$ and $\log \mathcal{L}_2$, and applying simplification, we obtain:

$$(\log \mathcal{L}_1 - \log \mathcal{L}_2) \propto \sum_{i=1}^{H} (d'_{i,1} - d'_{i,2})(d'_{i,1} + d'_{i,2} - 2d_i), \tag{23}$$

As illustrated in Fig. 10, the difference between $d'_{i,1}$ and $d'_{i,2}$ is extremely small at each individual baseline. Consequently, when the quantity of observed data $H$ is limited, the cumulative difference of $(d'_{i,1} - d'_{i,2})$ for $i=1,\ldots,H$, remains negligible. That is, $\mathbf{q}_1$ can be translated to various locations within the prior distribution, yet it still yields nearly identical likelihoods. Under these conditions, non-uniqueness of solutions occurs that results in a remarkably high level of uncertainty in the inversion results and significant deviations in the posterior means. However, if the quantity of deformation data is substantial, for instance, 50 and 100 data points, the cumulative differences of ($d'_{i,1}$–$d'_{i,2}$) for $i=1,…,H$, become more noticeable. This difference then distinguishes log $\mathcal{L}_1$ from log $\mathcal{L}_2$, subsequently affecting the posterior distribution. Consequently, the increasing volume of observed data gradually reduces the uncertain in the inversion results. This leads to the posterior distribution and posterior means converging toward the true values.

*3.3    Strategies for mitigating the uncertainty*

According to the above analysis, an effective strategy to mitigate inversion uncertainty is to increase the quantity of deformation data. Here, we will discuss an even more effective strategy.

As revealed, one crucial factor leading to non-uniqueness and a high level of uncertainty is the unknown uniform pressures. As illustrated in Fig. 11(a), the uniform pressures are converted into identical normal forces inside the tunnel lining due to the arch effect. In other words, if the normal force on a section of the structure can be identified, the unknown uniform pressures can be determined accordingly. Thus, it is believed that combining information from both the normal force and deformation data can mitigate non-uniqueness and reduce uncertainty. To verify this, we set up another 6 cases. Cases A1–F1 are configured identically to cases A–F, with the sole exception being the inclusion of a normal force measurement. Specifically, cases A1–F1 contain the same deformation data as cases A–F, respectively, in addition to the newly acquired normal force data at the tunnel crown (see Fig. 11a) in each case.

Incorporating normal force data, denoted as $N$, necessitates a slight adjustment to the likelihood function (Eq. 10):

$$p(\mathbf{d}, N | \mathbf{q}) = p(\mathbf{e_d}, \mathbf{e}_N), \qquad (24)$$

where, $\mathbf{e_d}$ and $\mathbf{e}_N$ signify the measurement errors associated with deformation and normal force, respectively. Given that deformation and normal force measurements typically employ distinct instruments, it is assumed that $\mathbf{e_d}$ and $\mathbf{e}_N$, follow independent

Gaussian distributions. Thus, we have:

$$p(\mathbf{d}, N | \mathbf{q}) = p(\mathbf{e_d} | \mathbf{q}) \cdot p(\mathbf{e}_N | \mathbf{q})$$
$$= p(\mathbf{e_d} | \mathbf{q}) \cdot \frac{1}{(2\pi\sigma_{e,N}^2)^{H_N/2}} \exp\{-\frac{[g_N(\mathbf{q}) - N]^T[g_N(\mathbf{q}) - N]}{2\sigma_{e,N}^2}\}, \quad (25)$$

where, $p(\mathbf{e_d}|\mathbf{q})$ remains consistent with Eq. (10), albeit annotated with a subscript to differentiate from the error terms associated with normal force observations; $g_N(\mathbf{q})$ calculates the internal normal force, incorporating a subsequent post-processing phase applied to the forward model $g(\mathbf{q})$; $H_N$ is the number of normal force data, which equals 1 in all cases A1–F1; $N$ is the observed normal force data at the tunnel crown, generated using the same process as that in Fig. 4, and measures 1623.6 kN for all 6 cases; $\sigma_{e,N}$ is the estimated standard deviation of normal force measurement errors, assumed to be approximately 1% of the measurement, corresponding to 15 kN in these cases.

All other conditions in cases A1–F1 remain the same as those in cases A–F. The inversion results are given in Figs. 12(a)–(f), respectively. Notably, these results once again affirm that PM effectively addresses the ill-conditioning issue encountered by OS. In addition, PM fits excellently with TP in case F1, with an index of agreement value $IA_{PM}=0.97$ and $RMSE_{PM}=31.8$ kPa, indicating that the number of parameters (22) is set densely enough to cover the complex true pressures.

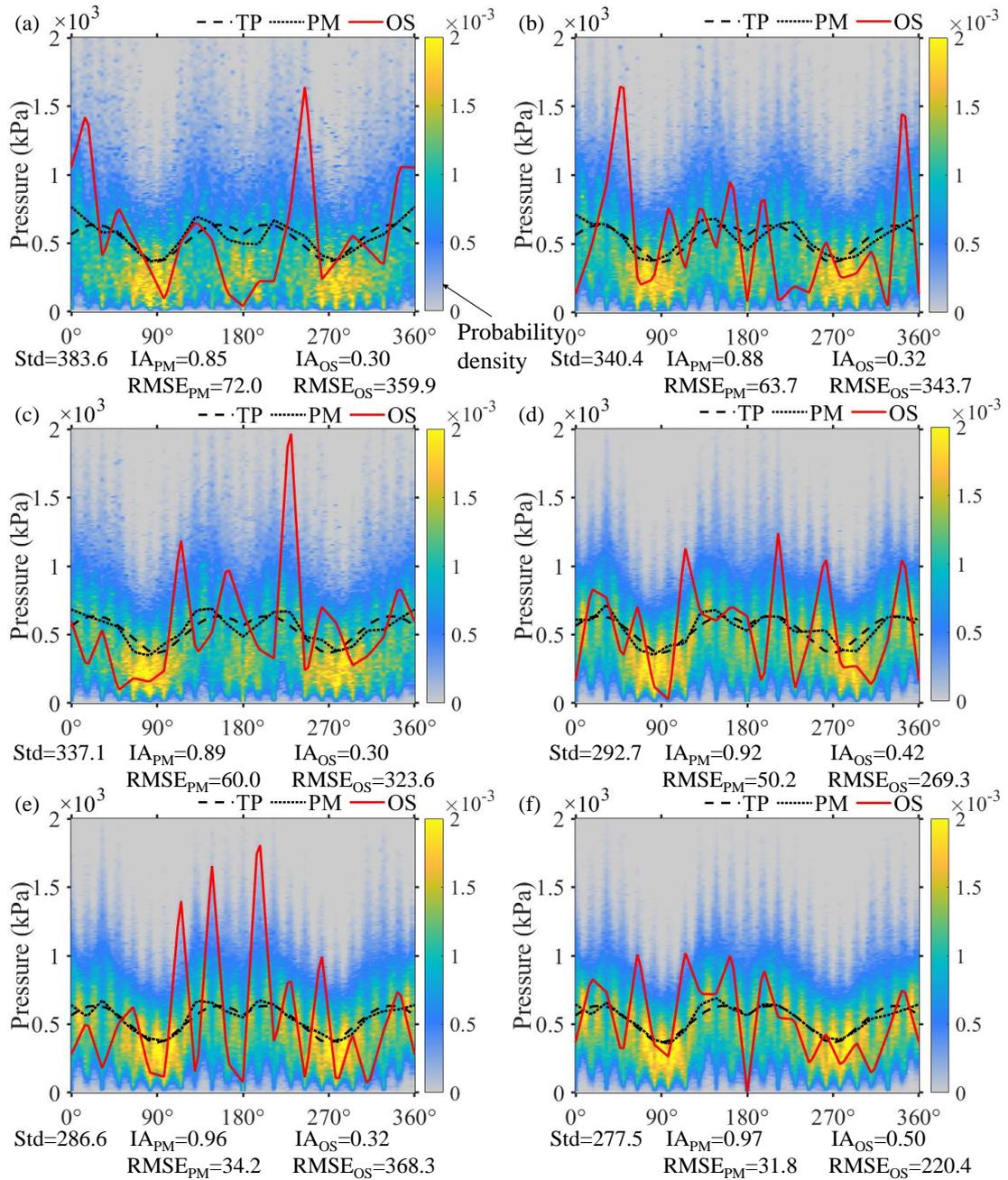

**Fig. 12.** Inversion results for: (a) case A1 (2 convergence data); (b) case B1 (5 convergence data); (c) Case C1 (10 convergence data); (d) case D1 (25 convergence data); (e) case E1 (50 convergence data); (f) case F1 (100 convergence data). (Note: all cases include one additional normal force observation at the tunnel crown; TP= true pressures; PM=posterior mean; OS=optimal solution obtained by deterministic inversion; the y-axis are rescaled from 0 to 2000 kPa to provide a clearer result).

Most importantly, upon comparing Figs. 8 and Figs. 12, it becomes evident that uncertainty significantly get reduced with the incorporation of normal force data. For instance, in the case with two convergence data points (Case A1), when the normal force data is incorporated, both the posterior distribution and posterior mean converge toward and fit satisfactorily with TP, resulting in an index of agreement value of $IA_{PM}$=0.85 and $RMSE_{PM}$=72.0 kPa. Furthermore, as the number of deformation data points increases, uncertainty gradually diminishes, and the fit of the posterior mean improves on the whole, reaching an $IA_{PM}$ value of 0.97 and $RMSE_{PM}$=31.8 kPa in the case with 100 convergence data points.

This trend is more clearly illustrated in Fig. 13, where uncertainties and $IA_{PM}$ values are summarized for both cases A–F and cases A1–F1. Measurement errors may introduce disturbances, resulting in non-smooth developments of IA and Std, particularly when data quantities are limited. However, the overall trend is clear. Specifically, the uncertainty in the results (Std) deceases rapidly with an increase in deformation data, tending to gradually converge when the deformation data exceeds 20. Notably, the addition of normal force data leads to a sharp reduction in uncertainty, particularly when the data is limited, such as with only 2 or 5 data points. Similarly, the fit of the posterior mean improves as the quantity of deformation data increases, and this improvement is especially pronounced when normal force data is incorporated.

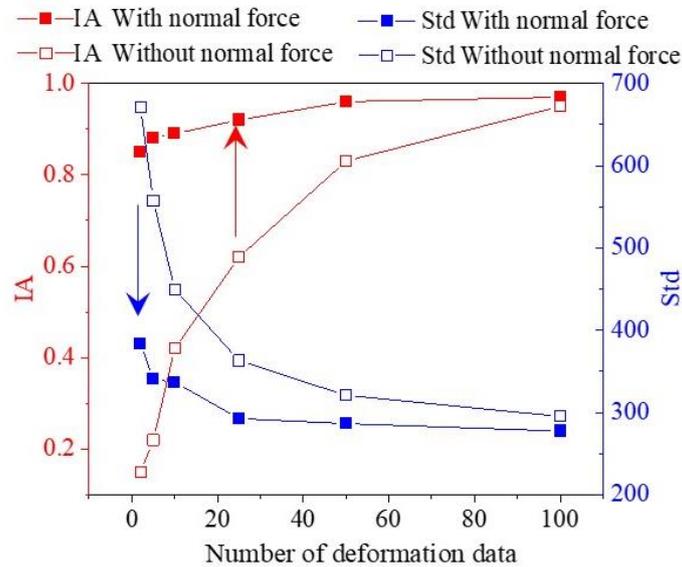

**Fig. 13.** Summation of fitness and uncertainty of the inversion results for both cases A–F and cases A1–F1.

In summary, the primary source of uncertainty arises from the intrinsic non-uniqueness of the mechanical system. To address this uncertainty, two strategies can be considered: increasing the quantity of deformation data and incorporating normal force data into the deformation dataset. Notably, the latter strategy proves to be much more effective. The following sections will explore practical application and experimental testing of these findings.

## 4 Application and verification
### 4.1 Preliminaries

A numerical case study is conducted for necessary application and verification. As shown in Fig. 14, a tunnel lining is embedded in a soft soil stratum. The abrupt and unexpected deposition of dumped soils were loaded onto the ground surface, leading to significant deformation on both the ground and structures. The engineers sought to assess the load conditions acting on the tunnel linings by measuring convergence deformations inside the tunnel. Accordingly, the objective is to invert the unknown pressures applied to the tunnel lining.

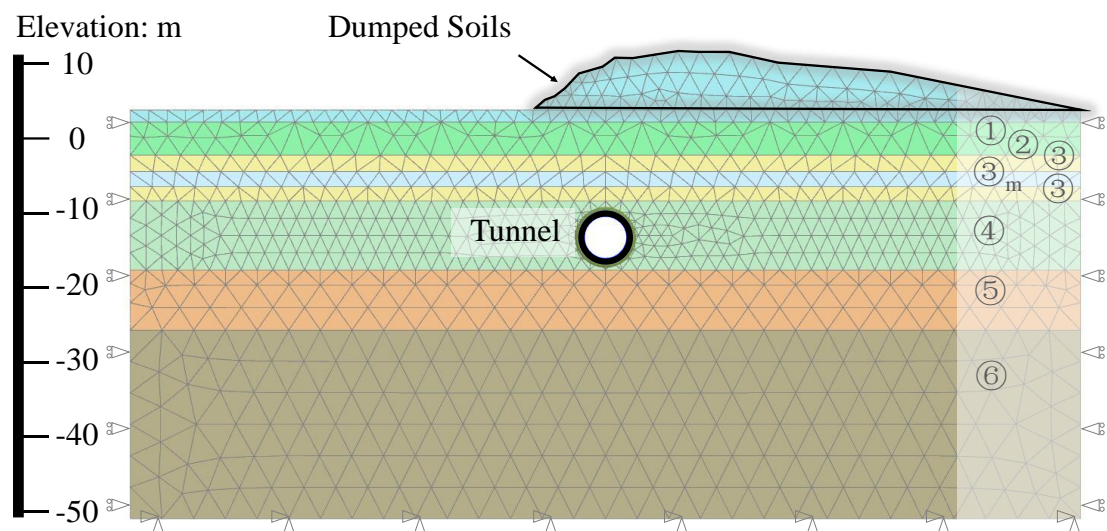

**Fig. 14.** The numerical case along with corresponding geological units and the FEM mesh.

This case was simulated using the PLAXIS 2-D software package. The initial cover depth of the tunnel is 9.6 m. Subsequently, dumped soil is loaded onto the ground surface with a maximum height of 7.8 m. To simplify the simulation, the tunnel lining is considered to be continuous and linear elastic, but a reduction factor, i.e., $\eta=0.26$ [27], is applied to account for the reduced rigidity due to joints. The tunnel ring has a diameter of 6.2 m, a width of 1 m, a thickness of 0.35 m, and a Young's modulus of $3.5 \times 10^7$ kPa. The stress-strain behavior of the soil materials is described using the Mohr-Coulomb model, with detailed parameters listed in Table 1. Additionally, a reduction factor of $R_{inter}$=0.5 is applied to reduce the strength and stiffness of the soil-structure interface [40]. Detailed geological units and the finite element mesh are illustrated in Fig. 14.

Table 1. Parameter settings for this numerical model

| Number | Stratum | $\gamma$ /(kN•m$^{-3}$) | $c$ /kPa | $\varphi$ /° | $E_s$ /MPa | $\mu$ |
|---|---|---|---|---|---|---|
| ① | Fill | 19.0 | 15.0 | 20.2 | 4.3 | 0.33 |
| ② | Silty lay | 18.4 | 17.0 | 21.5 | 5.4 | 0.32 |
| ③ | Muddy silty clay | 17.6 | 9.0 | 16.5 | 3.4 | 0.35 |
| ③$_m$ | Sandy silt | 18.1 | 17.0 | 14.0 | 4.3 | 0.34 |
| ④ | Muddy clay | 16.8 | 13.0 | 10.5 | 2.3 | 0.37 |
| ⑤ | Clay | 18.5 | 15.0 | 15.0 | 8.6 | 0.31 |
| ⑥ | Mudstone | 20.0 | 1.0 | 40.0 | 200.0 | 0.20 |

Note: $\gamma$ is the unite weight; $c$ is the cohesion; $\varphi$ is the friction angle; $E_s$ is the compressive modulus; $\mu$ is the Poisson's ratio

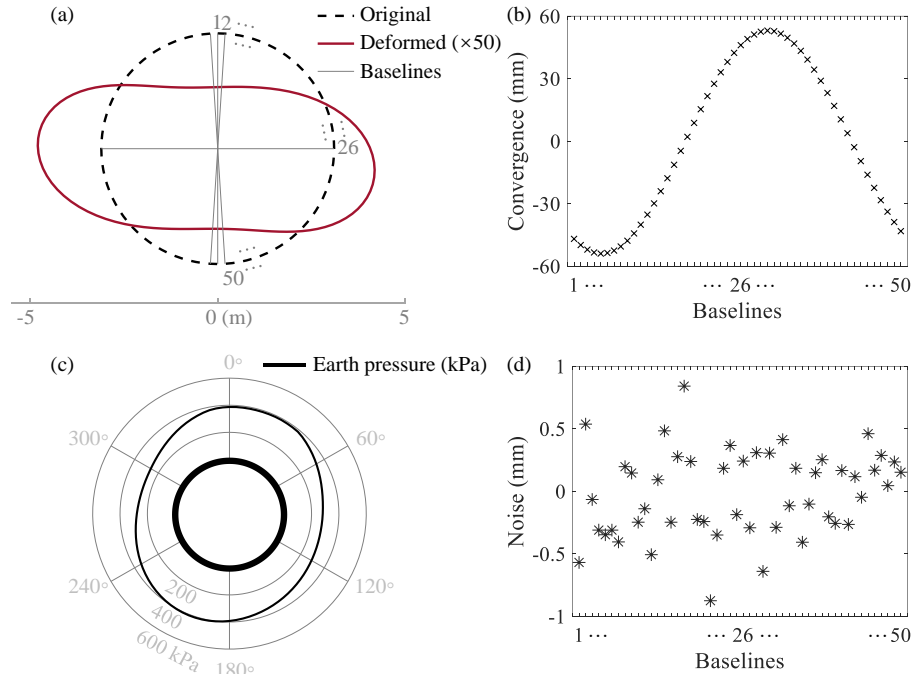

**Fig. 15.** The numerical example: (a) deformation profile of the tunnel produced by the numerical simulation; (b) the observed convergence deformation on the defiend baselines; (c) the actual pressures applied on the tunnel lining; (d) the Gaussian noise to simulate measurement errors.

The simulation can be roughly divided into two steps. In the first step, the soil is excavated, and the tunnel lining is generated. In the subsequent step, the dumped soil is activated to be loaded onto the ground surface. The final deformation profile of the tunnel ring is calculated and presented in Fig. 15(a). Subsequently, 50 evenly distributed baselines, as shown in this figure, are defined, and the convergence deformations on these baselines are observed and displayed in Fig. 15(b). In addition, the normal force at the tunnel crown is measured to be 880.7 kN, and will also be used in the inversion process. To evaluate the inversion accuracy, the true pressures on the tunnel lining are extracted from the numerical software, presented in Fig. 15(c). To simulate the measurement errors, we introduced random noise into the convergence data, which is given in Fig. 15(d).

Consequently, the objective here is inversion of the earth pressures based on the noise-corrupted structural responses. Six different cases with different number and kind of observed data were considered. Specifically, case A2, C2, and E2 comprise 10, 25, and 50 evenly distributed convergence data, respectively; while cases B2, D2, and F2 are configured identically to cases A2, C2, and F2, respectively, with the sole exception being the inclusion of a normal force measurement at the tunnel crawn.

Regarding the inversion process, we evenly distributed 22 unknown parameters across the structural domain, i.e., $\mathbf{q}=(q_1, …, q_{22})^T$. The prior distribution of these parameters is set as a uniform distribution $q_i \sim$ Uniform $(0,3000)$ ($i$=1,…,22). $\sigma_e$ in the likelihood function is determined to be 1 mm. The forward model in the likelihood function is constructed following the process outlined in section 2.2.3. It is noteworthy for the following points: i) no joints need to be considered in the forward model, as the tunnel is treated as continuous in the numerical case; ii) the soil reaction stiffness is set to be linear at 1000 kN/m$^3$, a topic that will be discussed in section 4.3.2. iii) an additional step to eliminate the reaction pressures of soil springs is required in this case, as we are inverting the final pressures acting on the tunnel lining. The number of the Markov Chain components is set to be 44, and the iteration times are set to be 100000.

*4.2 Results and verification*

Inversion results for cases A2–F2 are presented in Figs. 16(a)–(f), respectively. As shown in the figures, when the deformation data is limited, and no normal force data are incorporated (case A2), the uncertainty of the inversion results is extremely high.

The high probability (light color) zone does not appear to cover the true pressures TP, and PM also deviates significantly from TP. However, despite fluctuations, the shape of PM seems to be somewhat similar to TP. With an increase in deformation data (cases C2 and E2), the uncertainty diminishes, and the posterior distribution and PM tends to gradually converge towards the true pressures.

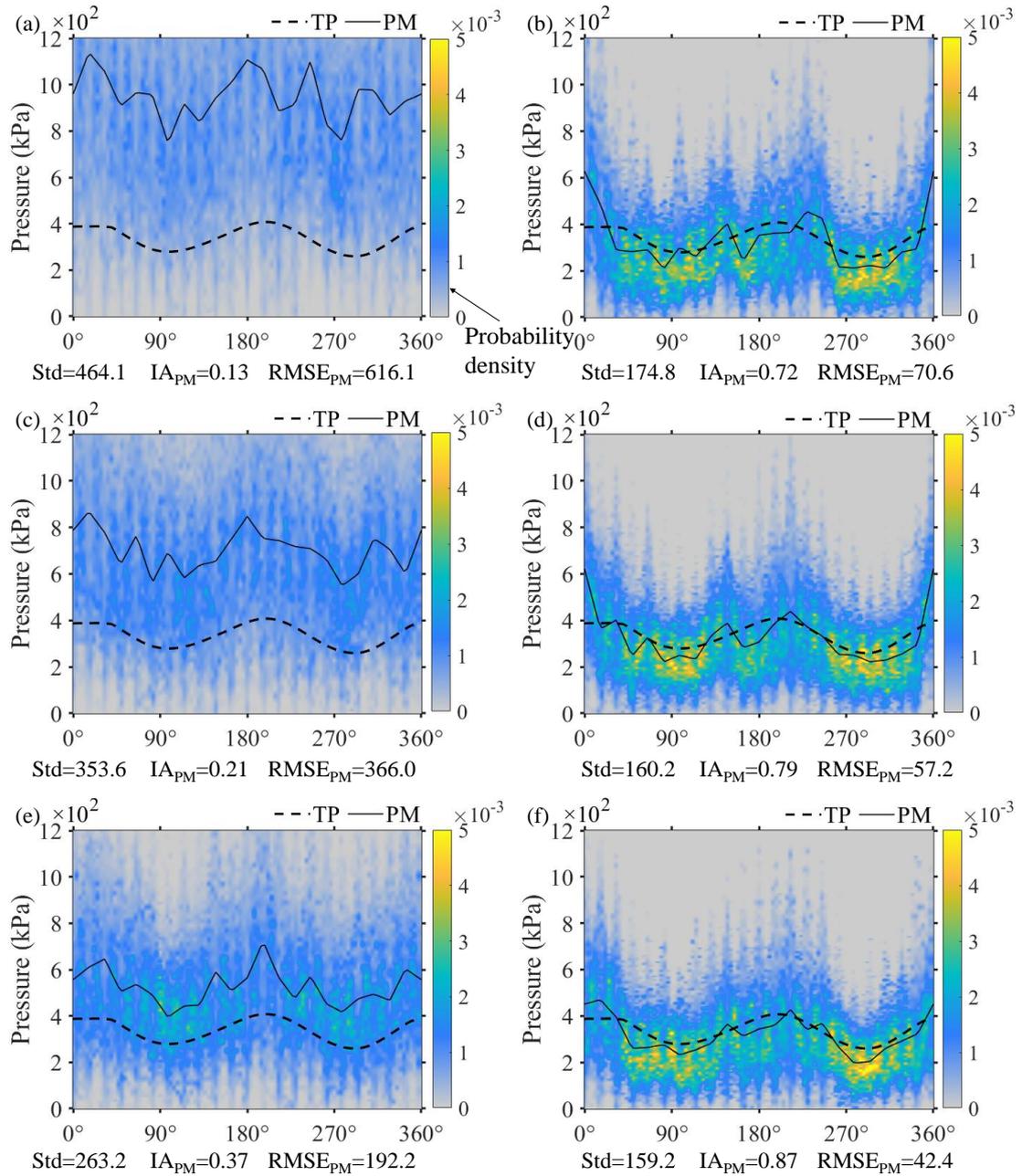

**Fig. 16.** Inversion results for: (a) case A2 (10 convergence data); (b) case B2 (10 convergence data with 1 additional normal force observation); (c) Case C2 (25 convergence data); (d) case D2 (25 convergence data with 1 additional normal force observation); (e) case E2 (50 convergence data); (f) case F2 (50 convergence data with 1 additional normal force observation). (Note: TP= true pressures; PM=posterior mean; the y-axis are rescaled from 0 to 1200 kPa to provide a clearer result)

By comparing cases A1 and B1, it is evident that when a normal force data is involved, uncertainty in the results is reduced substantially, with the posterior distribution and PM fitting well with TP. In addition, with the increase of deformation data (cases D2 and F2), the inversion results show continuous improvement. All these results corroborate the effectiveness of the findings presented in section 3.

Simultaneously, PM in case F2 fits satisfactorily with TP with an IA value of 0.87 and RMSE value of 42.4 kPa. This satisfactory agreement suggests that the presented Bayesian approach and corresponding findings hold significant promise for application in engineering. Nevertheless, further discussion is warranted before confidently applying this approach in practice, which will be detailed in the subsequent section.

*4.3   Discussion*

*4.3.1   Influence of measurement errors*

Measurement errors are widely acknowledged as a source of uncertainty that can influence inversion outcomes. To evaluate this impact, additional cases are established. Cases F3 and F4 are configured identically to case F2, but with different levels of measurement errors. Specifically, case F3 introduces Gaussian noise, following a distribution $\mathcal{N}(0,1^2)$, as presented in Fig. 17(c). Case F4 introduces Gaussian noise with a distribution $\mathcal{N}(0,2^2)$, as shown in Fig. 17(e). Notably, the maximum absolute error approximates 1 mm for case F2, which then increases to 3 mm and 5 mm for cases F3 and F4, respectively.

The inversion results for cases F3 and F4 are presented in Figs. 17(d) and (f), respectively. Figs. 17(a) and (b) display the inputs and results from case F2 for comparative purposes. It is evident that increasing noise levels result in a deterioration of the fit between the PM and TP. In case F3, the PM demonstrates reduced smoothness compared to case F2, with an IA of 0.78 and a RMSE of 60.0 kPa. As the noise level increases in case F4, the PM exhibits substantial fluctuations, evidenced by an IA of 0.69 and a RMSE of 77.1 kPa, indicating suboptimal inversion outcomes.

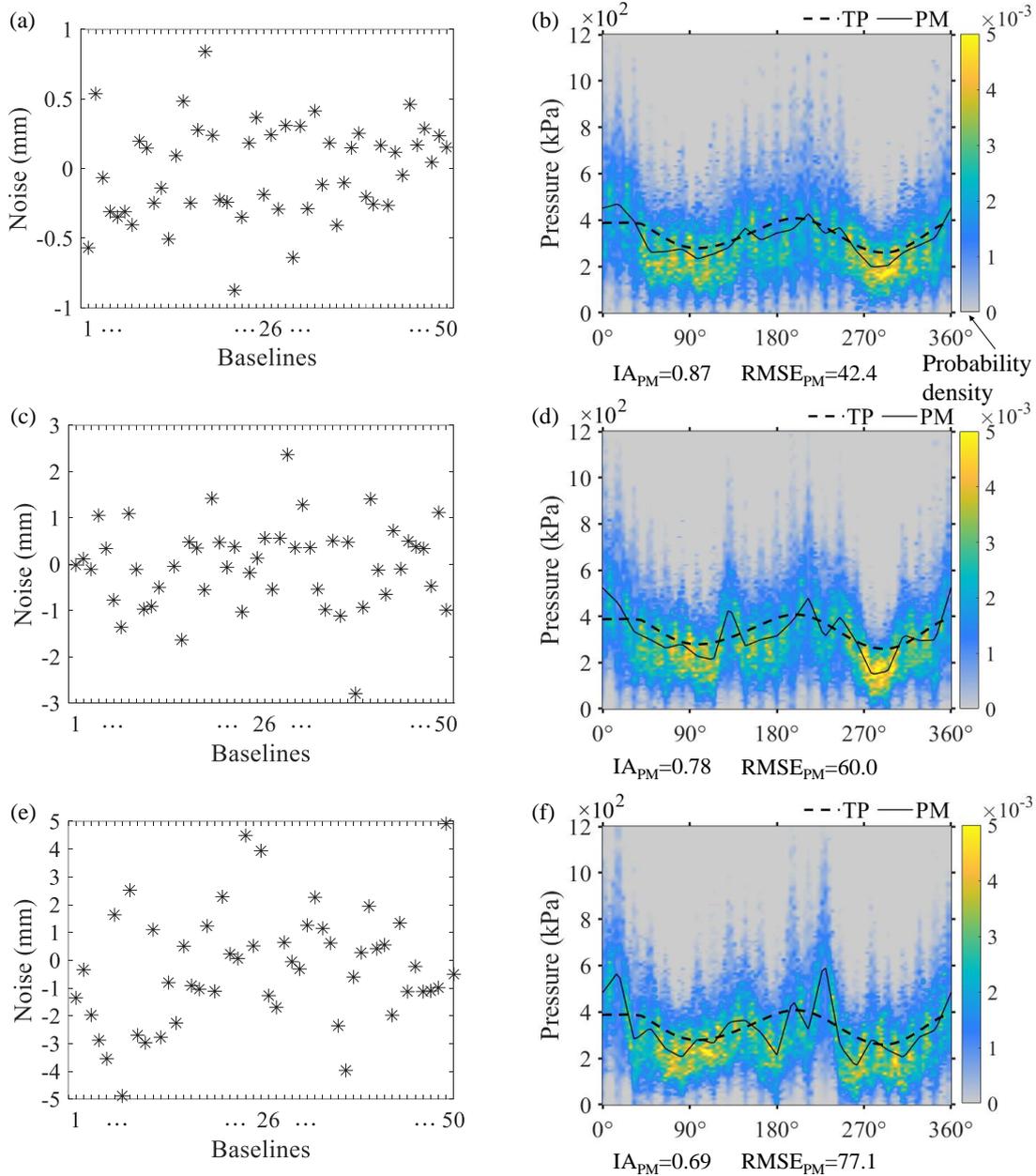

**Fig. 17.** The measurement errors and inversion results: (a) random noise for case F2; (b) inversion results for case F2; (c) random noise for case F3; (d) inversion results for case F3; (e) random noise for case F4; (d) inversion results for case F4. (Note: TP= true pressures; PM=posterior mean; the y-axis are rescaled from 0 to 1200 kPa to provide a clearer result)

Although measurement errors in engineering practices typically adhere to a zero-mean Gaussian distribution, it is important to acknowledge that the specific errors in each individual measurement may vary, even when following the same distribution. To ensure robustness in testing, cases F2–F4 are each repeated 10 times. During each repetition, the noises for the three cases continue to be randomly generated from $\mathcal{N}(0,1/3^2)$, $\mathcal{N}(0,1^2)$, and $\mathcal{N}(0,2^2)$, respectively, with verifications confirming that the maximum absolute error remains within 1 mm, 3 mm, and 5 mm, respectively. The results of the 10 repetitions for cases F2–F4 are summarized in Fig. 18. As demonstrated, random errors contribute to variability in results. However, the deterioration of results with increasing noise level is evident, with the average IA decreasing from 0.85 to 0.75 to 0.67 and the average RMSE increasing from 45.2 to 64.5 to 82.4 kPa across cases F2, F3, and F4, respectively. Consequently, in engineering applications, it is crucial to minimize measurement noise to ensure optimal inversion outcomes. The satisfactory results in case F2 suggest that maximum absolute errors should be maintained within [−1, 1] mm as much as possible.

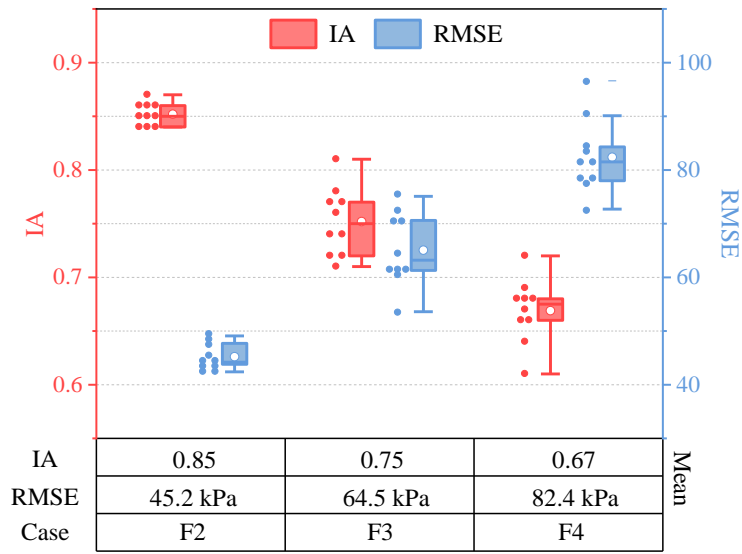

**Fig. 18.** Summary of results from 10 repetitions of cases F2, F3, and F4.

### *4.3.2 Role of the soil springs*

The strategy to address asymmetric earth pressures, detailed in Section 2.2.3, is informed by inherent soil-structure interaction. When environmental disruptions induce rigid body displacement in a tunnel lining, the subsequent interaction between soil and structure counterbalances the asymmetric component of the disruption, restoring stability of the lining. At this point, isolating the lining as the study subject, the applied actual pressures (referred to as net pressures) should systematically maintain stability, controlling relative deformations independently of rigid body displacement.

Informed from engineering practice, it is hypothesized that the relative deformations are primarily governed by the net pressures, while the rigid body displacement is influenced by the soil reaction pressures. Considering the data for inversion is the tunnel convergence data, which reflects relative deformations and are uncorrelated with rigid body displacement, we propose a conjecture: the settings and mechanical behavior of soil springs minimally impact inversion results, i.e., the net pressures. In other words, although higher soil spring stiffness leads to increased reaction pressures, total pressures on the beam-foundation structure also rise, thereby maintaining net pressures constant to match the observed convergence deformations.

The inversion results in case F2 provide partial support for this conjecture. Although the soil stiffness is modeled as linear at 1000 kN/m³—a value that does not accurately represent the properties of muddy clay—the inversion results are nonetheless satisfactory. For further exploration, two additional cases, F5 and F6, are conducted, configured identically to case F2, except for variations in soil reaction stiffness. In case F5, the stiffness is set at 100 kN/m³, while in case F6, it is set at 2000 kN/m³. As detailed in section 4.3.1, case F2 is repeated 10 times. The outcomes for the first implementation, i.e., with measurement errors given in Fig. 17a, are presented in Figs. 19. These results indicate that the results for cases F2, F5, and F6 are remarkably similar, with IA values of 0.87, 0.86, and 0.86; and RMSE values of 42.4, 43.4, and 43.1 kPa respectively. These minor differences may stem from the inherent randomness of the Markov Chain [31–32]. Specifically, according to Law of Large Numbers, the distribution of the chain samples will converge precisely to the target distribution as the chain length becomes infinite. However, in practical engineering, only a limited number of iterations can be performed on a chain (100000 in this case), which may lead to slight variations in the results as demonstrated.

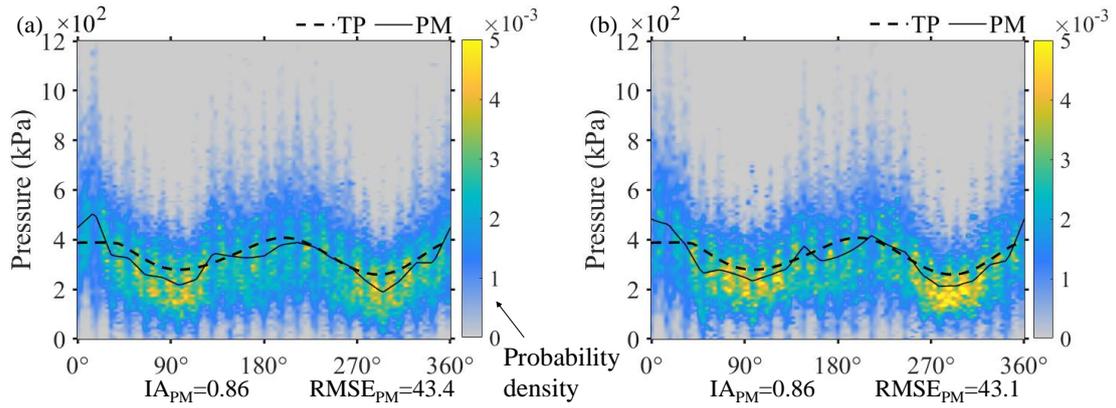

**Fig. 19.** Inversion results for first implementation of the 10 repetitions: (a) case F5; (b) case F6. (Note: TP= true pressures; PM=posterior mean; the y-axis are rescaled from 0 to 1200 kPa to provide a clearer result)

Similarly, for robustness testing, Cases F5 and F6 underwent 10 repetitions with the same input data as the 10 repetitions in Case F2. The results are summarized in Fig. 20. As shown, the average IA for all three cases stabilizes at 0.85, with average RMSE values of 45.2, 44.8, and 45.3 kPa for Cases F2, F5, and F6, respectively. These minor differences suggest that soil stiffness minimally affects the inversion results.

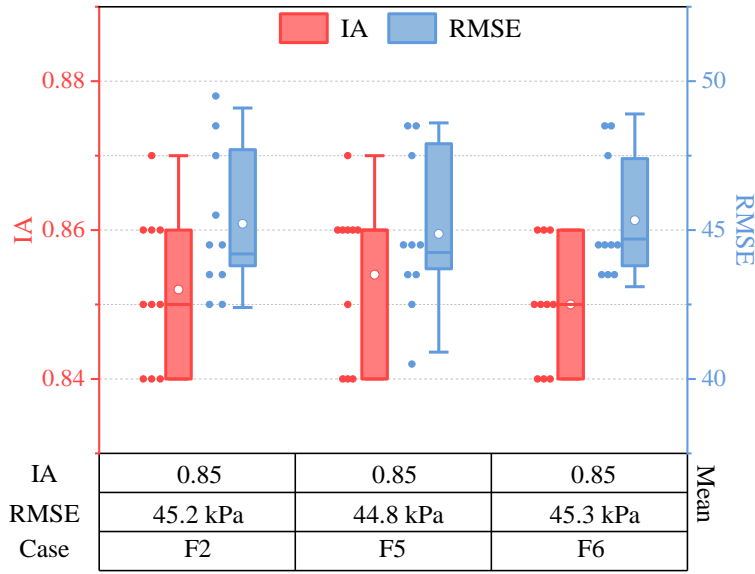

**Fig. 20.** Summary of results from 10 repetitions of cases F2, F5, and F6.

### 4.3.3 Future extensions

Both the illustrative example and the numerical case demonstrate that incorporating a measurement of normal force effectively reduces uncertainty in inversion processes and enhances accuracy. However, the widespread deployment of normal force sensors (e.g., reinforcement stress meter) along subway lines can be prohibitively expensive. Additionally, embedding sensors into already cast and constructed linings is technically unfeasible. In response, alternative observational techniques may serve as substitutes. For example, distributed fiber optic sensors or strain gauges could be installed on the internal tunnel surfaces to provide inferential data on normal forces [14, 41]. These installations could be restricted to linings that require health assessments, potentially reducing overall costs. Nonetheless, a potentially more economical and promising method is discussed subsequently.

Eq. (25) uses the difference between predicted and observed normal forces, $\{g_N(\mathbf{q})-N\}$, to construct the likelihood function, thus facilitating the incorporation of normal force into inversion analysis. Similarly, let $g_{N,M}(\mathbf{q})$ denote the function that predicts the normal force and moment vector $[N, M]$ for a specific section. It is important to recognize that tunnel linings requiring health assessments are typically accompanied by severe structural defects, such as tension cracks [42–44]. Development of these cracks is governed by $[N, M]$ acting on the concrete section [45–46]. For simplification, let $\mathbf{r}_c$ denote the observed crack characteristics, such as width, depth, and spacing; and let $\chi(.)$ denote the mapping from $[N, M]$ to predicted $\mathbf{r}_c$. Accordingly, if the crack details $\mathbf{r}_c$ can be observed and recorded as input data, it may be feasible to use $\{\chi([N,M])-\mathbf{r}_c\}$ as an alternative to $\{g_N(\mathbf{q})-N\}$ for constructing the likelihood, thereby providing insights into the normal forces within a segment. Notably, this approach necessitates careful establishment of $\chi(.)$ for segmental linings, warranting further studies in the future.

It is worth noting that all the forward models presented in this paper exhibit linear elastic behavior. However, the mechanical response of the tunnel lining in practical engineering is generally nonlinear, especially when considering crack development. It is desirable to extend the forward model to a high-fidelity one. However, a significant challenge lies in the fact that considering nonlinearity necessitates an iterative solution for the forward model. This implies that the total iteration times will be the product of the number of nonlinear iterations for the forward model and Markov Chain iterations, which can be highly time-consuming. Taking the numerical case as an example, the forward model is linear-elastic, with a single iteration, while the Markov Chain undergoes 10000 iterations. The computation is performed on a desktop with a Ryzen 9, 12-Core, 3.8GHz Processor. It is found that 10000 times iteration of the Differential Evolution-Markov Chain (DE-MC) for 44 components consumes 887 seconds. Ideally, when introducing non-linearity, this time will be multiplied by the iteration times of the forward model. Therefore, enhancing the efficiency of samplers for Markov chains becomes crucial. Potential improvements may include the utilization of GPU parallelization techniques and the adoption of gradient-based MCMC methods [47–48].

Acknowledging that when dealing with non-linear forward models in complex engineering applications, the model error may not be neglected in the likelihood function. In these instances, errors include both model errors and measurement errors simultaneously. Although the Central Limit Theorem may still be reasonably applied to assume a Gaussian distributed likelihood, the estimated standard deviation of the errors is not solely determined by the precision of measurement instruments. In such situations, an advanced hierarchical Bayesian framework [22] can be employed to jointly estimate the standard deviation along with the unknown pressures.

## 5  Conclusions

A Bayesian approach is presented in this paper for identifying earth pressures on in-service tunnel linings using easily observed deformation data. The distinctive feature of this approach, compared to existing deterministic approaches, lies in its Bayesian framework, enabling the quantification of inversion uncertainties. By quantifying uncertainties in the results, valuable insights can be drawn to enhance the overall outcomes. The key findings are summarized as follows:

i) The posterior mean serves as a representative solution within the Bayesian framework. It has been demonstrated to effectively handle ill-conditioning features encountered in deterministic inversions when measurements are contaminated by random errors. This emphasizes the superior robustness of the Bayesian approach for pressure inversion in tunnel linings.

ii) Uncertainties in inversion results arise from two sources: the non-uniqueness of solution and random measurement errors. The major source of uncertainty is the former, primarily stemming from the uniform component in distributed pressures. This can be mitigated by employing two strategies: increasing deformation data or incorporating a normal force observation. The latter strategy proves to be much more effective. These insights provide practical principles for improving pressure inversion outcomes.

iii) Measurement errors can introduce uncertainties and adversely affect inversion results. To ensure satisfactory outcomes, it is recommended to maintain deformation measurement accuracy within the range of $[-1, 1]$ mm whenever possible. This standard may serve as a guideline for the required data quality necessary to conduct accurate earth pressure inversions.

It is important to note that all cases presented in this paper are characterized by linear-elastic mechanical behavior. Extending this approach to non-linear cases is a valuable endeavor. However, the major challenge lies in the convergence efficiency of the Markov Chain, which warrants further investigation. In addition, incorporating crack observations on in-service tunnel linings to provide information of internal normal force for improved inversion results is another aspect deserving future study.

**Appendix A: the embedded beam spring model**

This Appendix presents the FEM format of the embedded beam spring model:

$$g(\mathbf{q}) = \mathbf{K}^{-1}\mathbf{f}(\mathbf{I}(\theta)\mathbf{q}), \tag{A1}$$

where, the tunnel lining is discretized into many small elements, with a length of $L$ (200 elements in the illustration example and 100 elements in the numerical case); $\mathbf{K}$ is the global stiffness matrix assembled by the element stiffness matrix $\mathbf{k}_g^e$ using the criteria illustrated in Huebner et al. [49]; $\mathbf{f}$ is also assembled by the nodal forces vector $\mathbf{f}_g^e$. $\mathbf{k}_g^e$ and $\mathbf{f}_g^e$ are defined in the global coordinate system and transformed via the following

equations:

$$\mathbf{k}_g^e = \mathbf{T}\mathbf{k}^e\mathbf{T}^T, \tag{A2}$$

$$\mathbf{f}_g^e = \mathbf{T}\mathbf{f}^e, \tag{A3}$$

where, $\mathbf{k}^e$ and $\mathbf{f}^e$ are the element stiffness matrix and force vector in the local coordinate system of element $e$, respectively. $\mathbf{T}$ is the transformation matrix of coordinates:

$$\mathbf{T} = \begin{bmatrix} \cos(\theta_e) & -\sin(\theta_e) & 0 & 0 & 0 & 0 \\ \sin(\theta_e) & \cos(\theta_e) & 0 & 0 & 0 & 0 \\ 0 & 0 & 1 & 0 & 0 & 0 \\ 0 & 0 & 0 & \cos(\theta_e) & -\sin(\theta_e) & 0 \\ 0 & 0 & 0 & \sin(\theta_e) & \cos(\theta_e) & 0 \\ 0 & 0 & 0 & 0 & 0 & 1 \end{bmatrix}, \tag{A4}$$

where, $\theta_e$ is the coordinate of element $e$ in the polar system of the tunnel as shown in Fig. 3(a).

$\mathbf{k}^e$ comprises the stiffness matrix of the beam $\mathbf{k}^e_b$ and the foundation (soil springs) $\mathbf{k}^e_f$:

$$\mathbf{k}^e = \mathbf{k}_b^e + \mathbf{k}_f^e, \tag{A5}$$

where,

$$\mathbf{k}_b^e = \begin{bmatrix} \dfrac{EA}{L} & 0 & 0 & -\dfrac{EA}{L} & 0 & 0 \\ 0 & \dfrac{12EI}{L^3} & \dfrac{6EI}{L^2} & 0 & -\dfrac{12EI}{L^3} & \dfrac{6EI}{L^2} \\ 0 & \dfrac{6EI}{L^2} & \dfrac{4EI}{L} & 0 & -\dfrac{6EI}{L^2} & \dfrac{2EI}{L} \\ -\dfrac{EA}{L} & 0 & 0 & \dfrac{EA}{L} & 0 & 0 \\ 0 & -\dfrac{12EI}{L^3} & -\dfrac{6EI}{L^2} & 0 & \dfrac{12EI}{L^3} & -\dfrac{6EI}{L^2} \\ 0 & \dfrac{6EI}{L^2} & \dfrac{2EI}{L} & 0 & -\dfrac{6EI}{L^2} & \dfrac{4EI}{L} \end{bmatrix}, \tag{A6}$$

$$\mathbf{k}_f^e = k_f \cdot \begin{bmatrix} 0 & 0 & 0 & 0 & 0 & 0 \\ 0 & \dfrac{13L}{35} & \dfrac{11L^2}{210} & 0 & \dfrac{9L}{70} & \dfrac{-13L^2}{420} \\ 0 & \dfrac{11L^2}{210} & \dfrac{L^3}{105} & 0 & \dfrac{13L^2}{420} & \dfrac{-L^3}{140} \\ 0 & 0 & 0 & 0 & 0 & 0 \\ 0 & \dfrac{9L}{70} & \dfrac{13L^2}{420} & 0 & \dfrac{13L}{35} & \dfrac{-11L^2}{210} \\ 0 & \dfrac{-13L^2}{420} & \dfrac{-L^3}{140} & 0 & \dfrac{-11L^2}{210} & \dfrac{L^3}{105} \end{bmatrix}, \tag{A7}$$

where, $k_f$ is the stiffness of the foundation.

For the elements located at the two sides of the joint, rotation spring should be accounted. This accounting adopts an adjusted matrix [50]:

$$\mathbf{k}_{b,j}^e = \mathbf{k}_b^e \mathbf{A}, \tag{A8}$$

where, $\mathbf{k}^e_{b,j}$ is the adjusted stiffness matrix when considering joints, and

$$\mathbf{A} = \begin{bmatrix} 1 & 0 & 0 & 0 & 0 & 0 \\ 0 & \dfrac{r_i r_j + 4r_j - 2r_i}{4 - r_i r_j} & \dfrac{2L r_i (r_j - 1)}{4 - r_i r_j} & & & \\ 0 & \dfrac{-6(r_j - r_i)}{L(4 - r_i r_j)} & \dfrac{-3r_i(r_j - 2)}{4 - r_i r_j} & & & \\ 0 & & & 1 & & \\ 0 & & & & \dfrac{r_i r_j + 4r_j - 2r_i}{4 - r_i r_j} & \dfrac{-2L r_j (r_i - 1)}{4 - r_i r_j} \\ 0 & & & & \dfrac{-6(r_j - r_i)}{L(4 - r_i r_j)} & \dfrac{-3r_j(r_i - 2)}{4 - r_i r_j} \end{bmatrix}, \tag{A9}$$

where, $r_i$ and $r_j$ are the rotation stiffness at the two ends of element $e$:

$$r = \dfrac{1}{1 + \dfrac{3EI}{k_\varphi L}}, \tag{A10}$$

where, $k_\varphi$ is the rotation stiffness of the joint. In a continuous beam without joints, $r$ equals 1.

$\mathbf{f}^e$ is equivalent to the distributed pressures $q(\theta)$, i.e., $\mathbf{I}(\theta)\mathbf{q}$, following the transformation rules of virtual work [51]. For simplification, define $\xi = \theta - \theta_e$, then:

$$\mathbf{f}^e = \begin{bmatrix} 0 & 0 & 0 & 0 & 0 & 0 \\ 0 & 1 & 0 & 0 & -\dfrac{3}{L^2} & \dfrac{2}{L^3} \\ 0 & 0 & 1 & 0 & -\dfrac{2}{L} & \dfrac{1}{L^2} \\ 0 & 0 & 0 & 0 & 0 & 0 \\ 0 & 0 & 0 & 0 & \dfrac{3}{L^2} & -\dfrac{2}{L^2} \\ 0 & 0 & 0 & 0 & -\dfrac{1}{L} & \dfrac{1}{L^2} \end{bmatrix} \begin{Bmatrix} 0 \\ F_{p0} \\ F_{p1} \\ 0 \\ F_{p2} \\ F_{p3} \end{Bmatrix}, \qquad (A11)$$

where,

$$F_{p0} = \int_{\theta_e}^{\theta_e+L} \mathbf{I}(\theta)\mathbf{q}\,d\theta \qquad F_{p1} = \int_{\theta_e}^{\theta_e+L} \mathbf{I}(\theta)\mathbf{q}\xi\,d\theta$$
$$F_{p2} = \int_{\theta_e}^{\theta_e+L} \mathbf{I}(\theta)\mathbf{q}\xi^2\,d\theta \qquad F_{p3} = \int_{\theta_e}^{\theta_e+L} \mathbf{I}(\theta)\mathbf{q}\xi^3\,d\theta \qquad (A12)$$

**Acknowledgements**

Z.T. was supported by Natural Science Foundation of China (Grant No. 51978523). A.L. was supported by the Engineering and Physical Sciences Research Council (EP/R034710/1). For the purpose of open access, the authors have applied a Creative Commons Attribution (CC BY) licence to any Author Accepted Manuscript version arising from this submission

**Data Availability Statement**

All data generated or used during the study are included in this paper. Code that supports the findings are available from Zhiyao.tian@bristol.ac.uk upon reasonable request.